\documentclass[twocolumn,preprintnumbers,amsmath,amssymb]{revtex4}
\usepackage{graphicx}
\usepackage{dcolumn}
\usepackage{bm}
\usepackage{hyperref} 
\usepackage{multirow} 
\usepackage{soul}
\newcommand\beq{\begin{equation}}
\newcommand\eeq{\end{equation}}
\newcommand\bea{\begin{eqnarray}}
\newcommand\eea{\end{eqnarray}}
\usepackage{wrapfig}
\newcommand{\nonum}{\nonumber}
\begin{document}
\title{\bf Bond-order wave phase, spin solitons and thermodynamics of a frustrated linear spin-1/2 Heisenberg antiferromagnet \\}
\author{\bf  Manoranjan Kuma${\rm \bf r^{1,2}}$, S. Ramasesh${\rm \bf a^{2}}$  and Z.G. Soo${\rm \bf s^1} $}
\address{\it {\rm  $ ^1 Department$ } of Chemistry, Princeton University, Princeton NJ 08544 \\}
\address{\it $ {\rm ^2Solid}$ State and Structural Chemistry Unit, Indian Institute of Science, Bangalore 560012, India,\\}
\date{\today}
\begin{abstract}
The linear spin-1/2 Heisenberg antiferromagnet with exchanges $J_1$, $J_2$ between first 
and second neighbors has a bond-order wave (BOW) phase that starts at the fluid-dimer 
transition at $J_2/J_1 = 0.2411$ and is particularly simple at $J_2/J_1 = 1/2$. 
The BOW phase has a doubly degenerate singlet ground state, broken inversion 
symmetry and a finite energy gap $E_m$ to the lowest triplet state. 
 The interval $0.4<J_2/J_1<1.0$ has large $E_m$ and small finite size corrections. 
Exact solutions are presented up to $N=28$ spins with either periodic or open 
boundary conditions and for thermodynamics up to $N=18$. The elementary excitations 
of the BOW phase with large $E_m$ are topological spin-1/2 solitons that separate BOWs with opposite 
phase in a regular array of spins. The molar spin susceptibility $\chi_M(T)$ is 
exponentially small for $T \ll E_m$ and increases nearly linearly with $T$ to a 
broad maximum. $J_1$, $J_2$ spin chains approximate the magnetic properties of the BOW 
phase of Hubbard-type models and provide a starting point for modeling alkali-TCNQ salts.
\vskip .4 true cm
\noindent PACS numbers: 71.10.Fd, 75.10.Pq, 75.60.Ch, 71.30.+h\\ 
\noindent Email: soos@princeton.edu 
\end{abstract}
\maketitle

\section{Introduction}
The extended Hubbard model (EHM) has competing on-site repulsion $U > 0$,
 intersite interaction $V > 0$ and electron transfer $t$ between neighbors in 
one dimension (1D) with evenly spaced sites. The half-filled case with 
one electron per site has several phases: a charge density wave (CDW) at $V > U/2$ 
with broken electron-hole symmetry and occupation numbers $n > 1$ on one 
sublattice, $n < 1$ on the other; a spin fluid phase at $V = 0$ as 
known from Hubbard models; and as proposed by Nakamura, \cite{rr1} 
a bond-order wave (BOW) phase with broken inversion symmetry between 
the CDW and the spin fluid phases when $t/U$ is sufficiently large for 
a continuous CDW transition. The BOW phase has long-range order and a 
finite magnetic gap $E_m$ to the lowest triplet excited state. 
Multiple theoretical approaches,\cite{r2,r3,r4,r5} primarily at $U \le 2t$ and $U=4t$, 
have confirmed a narrow BOW phase in the EHM. Other spin-independent 
potentials also support a BOW phase when the CDW transition is 
continuous \cite{r6}. The narrow BOW phase of Hubbard-type models presents major 
computational difficulties. 

In this paper, we consider the BOW phase of a familiar spin-1/2 chain 
with frustrated antiferromagnetic (AF) exchange \cite{r7}. The BOW phase becomes 
numerically accessible and can be demonstrated in finite systems. Although 
charge fluctuations are strictly excluded in spin chains, the BOW phase again 
illustrates broken inversion symmetry at sites, long-range order and finite $E_m$. 
The spin chain has AF exchange between first and second neighbors, 
\begin{eqnarray}
H(x)=J\sum_n ((1-x) \vec {s}_n.\vec {s}_{n+1}+x\vec {s}_n.\vec {s}_{n+2})
\label{eq1}
\end{eqnarray}
\noindent We consider the interval $0 \le x \le 1$ and set the total 
exchange $J = 1$ as the unit of energy. The $x = 0$ limit is a 
linear Heisenberg antiferromagnet (HAF). Second-neighbor exchange $J_2 = xJ$ 
for $x > 0$ opposes short-range antiferromagnet (AF) order and eventually induces a 
fluid-dimer phase transition that has been the focus of recent 
studies \cite{r8,r9,r10,r11,r12,r12p}. The $x=1$ limit gives two HAFs on 
the even and odd sublattice, respectively. White and Affleck \cite{r13p} considered 
 $J_2/J_1>1$ using field theory and the density matrix renormalization group (DMRG).\\

\begin{figure}[h]
\begin {center}
\hspace*{-0cm}{\includegraphics[width=8.5cm,height=3.0cm]{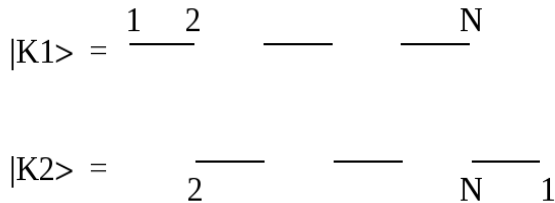}} \\
\caption{ Lines indicate singlet-paired spin on adjecent sites.}
\label{fig1p}
\end {center}
\end{figure}

Okamoto and Nomura \cite{r8} located the transition at $x_c= x_1/(1 - x_1) = 0.2411$, 
or $x_1 = 0.1943$ in our notation, where a magnetic gap $E_m$ opens. The ``dimer'' phase 
refers to the earlier observation of Majumdar and Ghosh (MG) \cite{r7} that, for an 
even number $N$ of spins and periodic boundary conditions (PBC), the exact ground state 
(gs) at $x_{MG} = 1/3$ $(J_2 = J_1/2)$ has singlet-paired spins on adjacent sites, 
just as in the  $ \rm Kekul\acute{e}$ diagrams shown in Fig. \ref{fig1p},
\begin{eqnarray}
|K1\rangle &=& (1,2)(3,4)...(N-1,N)  \nonum \\
|K2\rangle &=& (2,3)(4,5)...(N,1)  
\label{eq2}
\end{eqnarray}
\noindent $ \rm Kekul\acute{e}$ diagrams are particularly simple BOWs. They illustrate 
broken inversion symmetry, long-range order, double degeneracy and finite excitation 
energies \cite{r10} at the MG point. Since these are the defining features of 
a BOW phase, that is what we will call the ``dimer'' phase. The notation $H(x)$ 
is convenient for the $x = 1$ limit of HAFs on the even and odd sublattices. 
Now $x < 1$ describes interchain exchange that is frustrated because each 
spin is coupled to two neighbors of the other sublattice. With constant total 
exchange, the gs energy is highest when the spin chain is the most frustrated.\\

For open boundary conditions (OBC), $|K1 \rangle $ is the exact nondegenerate 
gs of $H(1/3)$. The chemical analogy is now to partial double and single 
bonds in linear polyenes or in polyacetylene. The BOW associated with $|K1 \rangle$ 
is well understood in {\it dimerized}  arrays whose elementary excitations are 
the topological solitons of the uncorrelated Su-Schrieffer-Heeger (SSH) model \cite{r13,r14}. 
Similar conclusions hold in correlated models of conjugated polymers \cite{r15} 
or ion-radical stacks \cite{r16}. Spin solitons in BOW systems at finite temperature 
separate $|K1 \rangle$ and $|K2 \rangle$ regions in infinite regular chains. 
The Peierls instability of Hubbard or spin chains is a separate topic that 
requires electron-phonon or spin-phonon coupling. Our discussion of $H(x)$ 
is limited to regular arrays with PBC or OBC.\\

\begin{figure}[h]
\begin {center}
\hspace*{-0cm}{\includegraphics[width=7.0cm,height=9.5cm,angle=-90]{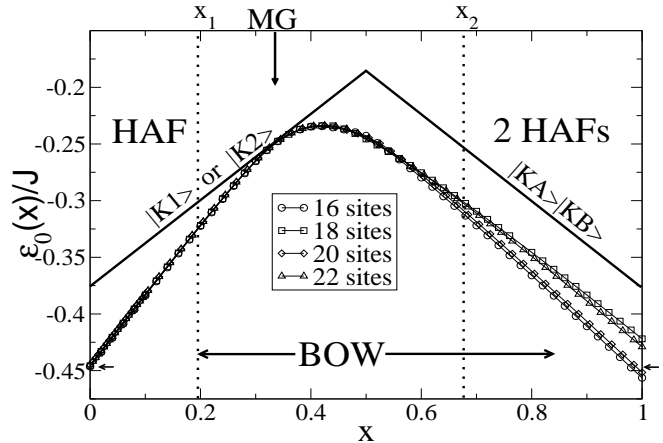}} \\
\caption{Ground state energy per site, $\epsilon_0(x)$, of the spin chain $H(x)$ in Eq. \ref{eq1} for $N$ 
sites with periodic boundary conditions (PBC). The BOW phase starts at the fluid-dimer transition at 
$x_1$; $x_2$ is the excited state crossover discussed in the text. 
The energy of $|K1\rangle$ or $|K2 \rangle$ in 
Eq. \ref{eq2} is exact at $x_{MG} =1/3$ and approximate for $x < 1/2$; 
the energy of a product of Kekul${\rm \acute{e}}$ 
diagrams $|KA\rangle|KB \rangle$  is approximate for $x > 1/2$. 
Arrows at $x = 0$ and 1 mark $\epsilon_0$ of the infinite chain.}       
\label{fig1}
\end {center}
\end{figure}

To introduce the principal features of $H(x)$, 
we show in Fig. \ref{fig1} the gs energy per site, $\epsilon_0(x)$, 
for intermediate $N \approx 20$ and PBC. Bonner and Fisher \cite{r23} 
found that $\epsilon_0(0)$ of the HAF converges as $\approx N^{-2}$ to 
the exact value, $-\rm ln2 + 1/4$, due to Hulthen \cite{r24} 
and denoted by arrows at $x = 0$ and 1. Convergence at $x = 1$ 
is for two HAFs of $N/2$ sites, from below when $N/2$ is even and 
from above \cite{r25} when $N/2$ is odd. The shape of $\epsilon_0(x)$ 
indicates different frustration at small $x$ for exchange $J_2$ in one HAF 
and at large $x$ for exchange $J_1$ between two HAFs. Frustration is greatest 
at the $\epsilon_0(x)$ maximum. The Hellmann-Feynman theorem gives

\begin{eqnarray}
\frac {\partial \epsilon_0(x)}{\partial x}&=&\frac{1}{N}\langle\psi_0(x)|\frac {\partial H}
{\partial x}|\psi_0(x)\rangle \nonum\\
&=&\frac{1}{N}\sum_n\langle(\vec {s}_n.\vec {s}_{n+2}-\vec {s}_n.\vec {s}_{n+1})\rangle 
\label{eq3}
\end{eqnarray}
\noindent The bond orders or spin correlation functions are equal at $\partial \epsilon_0/\partial x =0$ when there is 
equal choice for pairing with a first or second neighbor. The slope 
$\partial \epsilon_0/\partial x$ is steeper at $x = 0$ than at $x = 1$. 
First and second neighbor spin correlation functions are known \cite{r26} exactly at 
$x = 0$ and they add in Eq. \ref{eq3}. The slope at $x=1$ is just the first-neighbor 
correlation function.\\

The solid lines in Fig. \ref{fig1} are the energy of $|K1 \rangle$ or $|K2 \rangle$ 
for $x < 1/2$, exact at $x_{MG} = 1/3$, and of a product of 
$\rm Kekul\acute{e}$ diagrams of two HAFs for $x > 1/2$. The 
BOW phase that we characterize below starts at $x_1$. 
We found $x_2 \approx 2/3$ using the Okamoto-Nomura \cite{r8} treatment of $x_1$. 
The gap $E_m$ is exponentially small but finite for $x>x_2$, and the 
BOW phase extends to $x=1$ according to White and Affleck \cite{r13p}.\\

The order parameter $B(x)$ is the gs amplitude of the BOW, 
\begin{eqnarray}
B(x)=\frac{1}{N}\sum_n(-1)^n \langle \vec {s}_n.\vec {s}_{n+1}\rangle
\label{eq4}
\end{eqnarray} 
\noindent The two gs have $\pm B(x)$. It follows immediately that 
$B(1/3)=3/8$ for $|K1 \rangle$ or $|K2 \rangle$ in  Eq. \ref{eq2}. As shown below,
large $B(x)$ and $E_m(x)$ between $x=1/3$ and $x \approx 1/2$ make possible our detailed 
finite-N study of the BOW phase.\\

The complete basis of $H(x)$ has dimension $2^N$, 
since each spin-1/2 has two orientations, and the total spin $0 \le S \le N/2$ is conserved. 
Reflection $\sigma$ through sites corresponds to inversion symmetry $C_i$ at sites in 
the infinite chain. Valence bond (VB) methods \cite{r17,r18,r19,r20} are well suited 
for finite models that conserve $S$. A few states with any $S$ and $\sigma $ can be 
found exactly up to $N \approx 30$. The full spectrum is needed for thermodynamics and 
has been obtained \cite{r21} to $N = 16$, which we increase to $N = 18$. DMRG extends \cite{r22} thermodynamics 
to N = 64. The spin chain $H(x)$ benefits from the smaller basis compared to $4^N$ in Hubbard models 
with charge degrees of freedom. An even greater advantage may be the exact gs at 
the MG point for finite $N$. In contrast to the numerically difficult BOW phase 
of Hubbard-type models, the BOW phase of $H(x)$ is accessible to direct finite-$N$ modeling 
between $x \approx 1/3$ and $x \approx 1/2$.

 We characterize the gs properties of the BOW phase and its 
elementary excitations in Section II, including the magnetic gap $E_m(x)$, 
the  order parameter $B(x)$, excited states at the MG point, and the bond-order 
domain walls of spin solitons. The temperature dependence of the molar spin 
susceptibility  $\chi_M(T)$ and specific heat $C(T)$ are found in Second III. 
Following an activated regime that depends of $E_m$, $\chi_M(T)$ increases 
almost linearly with $T$ in the BOW phase, quite differently from an HAF 
or an EHM with $t \ll (U - V)$. The Discussion relates $H(x)$ to the EHM and 
to $\pi$-radical salts with $\chi_M(T$) nearly linear in $T$.

\section{Ground and low-energy states}

\begin{figure}
\begin {center}
\hspace*{-0cm}{\includegraphics[width=7.0cm,height=9.0cm,angle=-90]{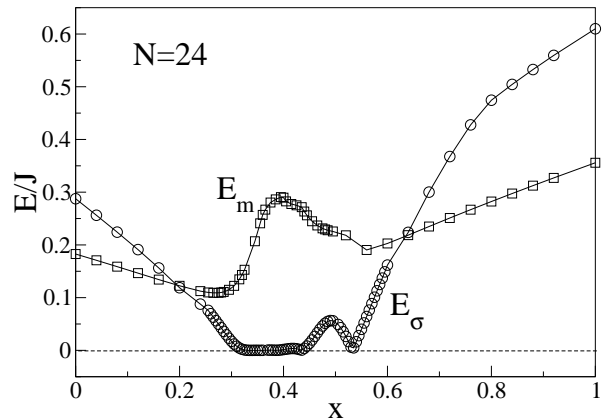}} \\
\caption{Finite-size effects on excitation energies of $H(x)$ for $N = 24$ with PBC. $E_m$ is the 
excitation to the lowest triplet and $E_{\sigma}$ to the lowest singlet with opposite 
inversion symmetry. The crossovers $x_1(24)$ and $x_2(24)$ are listed in Table \ref{tb1}.} 
\label{fig2}
\end {center}
\end{figure}

We use valence bond (VB) methods \cite{r19,r20} 
to solve $H(x)$ exactly for finite $N$ and either periodic or 
open boundary conditions in exact subspaces with fixed total $S$ 
and reflection $\sigma = \pm 1$ at sites. The gs 
is a singlet, $S = 0$, with either $ \sigma = 1$ or -1 
depending on $N$ and $x$. The linear combinations $|K1 \rangle \pm |K2 \rangle$ in Eq. \ref{eq2} transform as $\sigma = \pm 1$, 
even or odd under inversion in the infinite chain. The gs for other $x$ is 
a symmetry adapted linear combination of singlet VB diagrams $|k \rangle$. We define 
$E_{\sigma}(x)$ as the excitation energy to the lowest singlet with opposite $\sigma$ 
symmetry. Figure \ref{fig2} compares $E_{\sigma}(x)$ for $N = 24$ and PBC to the 
gap $E_m(x)$ to the lowest triplet. Finite-size effects are large at $x = 0$ for 
an HAF of 24 sites and about twice as large at $x=1$ for two HAFs of 12 sites, 
as expected when excitations energies go as $ \approx 1/N$. We obtained similar 
graphs of $E_{\sigma}(x)$ and $E_m(x)$ up to $N = 28$. There is no difference 
at small $x$ aside from $1/N$ effects. The excitations are qualitatively 
different at $x \approx 1$, however, when $N/2$ is even or odd. Even $N/2$ 
is required for proper comparison at $x = 0$ and 1. When $N/2$ is odd, 
the $x = 1$ limit corresponds to two HAFs with an odd number of spins and 
a doublet gs. Since the radical also has two-fold orbital degeneracy \cite{r25}, 
there are several gapless excitations in the $x = 1$ limit of no interchain exchange.\\ 

Okamoto and Nomura \cite{r8} identified the quantum transition at $x_1$ by finding 
$E_{\sigma}(x_1) = E_m(x_1)$ from $N$ = 10 to 24. They argued that an excitation 
crossover at finite $N$ is more accurate than extrapolation to find $E_m(x_1) = 0$. 
The slow variation of $x_1(N)$ and extrapolations made possible their accurate 
determination of $x_1$. Our results for $x_1(N)$ in Table 1 to $N$ = 24 agree with ref. 8. 
Previous work \cite{r8,r9,r10,r11,r12,r12p} focused on the fluid-dimer transition at $x_1$, 
while we are interested in the BOW phase with $x > x_1$. The same method yields $E_m = E_{\sigma}$ 
at $x_2(N)$ in Table \ref{tb1} for even and odd $N/2$. Finite-size effects are 
stronger because the chains are effectively half as long. A joint $1/N$ extrapolation of the 
two sequences returns $x_2 = 0.67 \pm 0.01$ ($J_2/J_1 = 2.03 \pm 0.03$). The $x_2$ 
crossover does not signify the termination of BOW phase, however, which extends \cite{r13p} to $x=1$.
We improved the accuracy of the DMRG algorithm \cite{r25p} to look at $x>x_2$ and find small but finite 
$E_m(x)$ and $B(x)$ up to $x=0.8$ $(J_2/J_1=4.0)$, beyond which even more accurate DMRG is required. 
We do not understand the different implication of $x_1$ and $x_2$ crossovers, but note that $H(x)$ also 
has a spiral phase \cite{r25pp} starting at $x_{MG}=1/3$ whose order parameter is twist angle. 
In the present work we focus on the BOW phase with large $B(x)$ and $E_m(x)$.\\ 

\begin{table}
\begin{center}
\caption {Crossing points $x_1(N)$ and $x_2(N)$ where $E_m(x) = E_{\sigma}(x)$ for $N$ sites and PBC.}
\begin{tabular}{cccc} \hline
$~~~~N~~~~$ & $ ~~~~x_1~~~~$ & $x_2$ ($N/2$  odd )  & $ x_2$ ($N/2$ even)  \\\hline
18 & 0.1949 &  0.5669 &\\
20& 0.1947 & & 0.6262 \\
22& 0.1947 &  0.5784 &\\
24& 0.1946 & & 0.6368 \\
26& 0.1946 &  0.5885&\\
28& 0.1944 & &0.6421 \\\hline
$\infty$ &  0.1943 \footnote{ ref.8} & 0.68 & 0.66\\\hline
\end{tabular}
\label{tb1}
\end{center}
\end{table}

\begin{figure}
\begin {center}
\hspace*{-0cm}{\includegraphics[width=7.0cm,height=9.0cm,angle=-90]{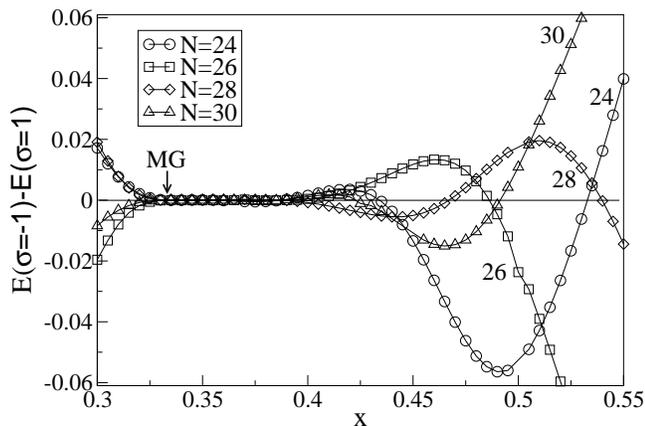}} \\
\caption{Finite-size effects on the energy the lowest singlets of $H(x)$ 
with opposite inversion symmetry in the BOW phase. The energies are equal 
at $x = 1/3$ without crossing. Crossings at $x > 1/3$ are governed by 
the ground-state symmetry at $x = 0$ and 1 as discussed in the text.}
\label{fig3}
\end {center}
\end{figure}

The BOW phase of the extended system has degenerate gs in the $\sigma = \pm 1$ sectors and 
hence $E_{\sigma}(x) = 0$. Finite-size effects are extraordinarily small for $0.3 < x < 0.5$ 
where $E_0$ is close to $-N/4$. The difference in {\it total} energy, $E_0(x,1)-E_0(x,-1)$, for $\sigma = \pm 1$ 
is shown in Fig. \ref{fig3} up to $N = 30$. The absolute gs for $x < 1/3$ is $E_0(-1)$ for $N = 4p$ 
and $E_0(1)$ for $N = 4p+2$. The $E_0(x,\pm 1)$ curves are degenerate at $x = 1/3$ 
without crossing. They cross for $x > 1/3$ and the number of crossings depends on $N$. 
When $N = 4p$ and $p$ is even, the gs at $x = 1$ also transforms as $\sigma = -1$; there is 
an even number of crossings for $N = 24$ or 16. The $E_{\sigma}$ bump in Fig. \ref{fig2} at 
$x = 0.5$ is due to two crossings. For $N = 4p$ and odd $p$, the $x = 1$ gs has $\sigma = 1$ 
symmetry that requires an odd number of crossings for $N =28$ or 20. The gs for $N = 4p+2$ 
has $ \sigma = 1$ symmetry for $x < 1/3$. There is an even number of crossings up to $x = 0.60$ for $N = 26$ 
or 18 (even $p$) and an odd number for N = 30 or 22 (odd $p$). Degeneracy without crossing at the MG point 
and subsequent symmetry crossovers for finite $N$ are the principal reasons for remarkably small $E_{\sigma}(x)$ 
in this interval.
 
\begin{figure}
\begin {center}
\hspace*{-0cm}{\includegraphics[width=7.0cm,height=9.0cm,angle=-90]{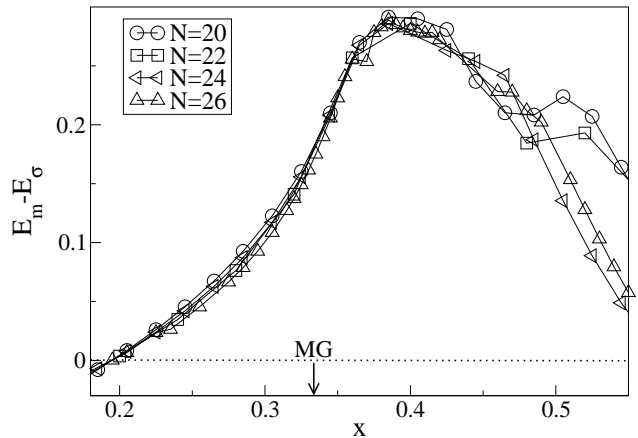}} \\
\caption{ Excitation energy $E_m(x)$ to the lowest triplet of $H(x)$ for finite $N$ and PBC in the 
BOW phase. The infinite chain has $E_{\sigma}(x) = 0$ and a doubly degenerate ground state.} 
\label{fig4}
\end {center}
\end{figure}

Since $E_{\sigma}(N) > 0$ is due to finite-size effects in the BOW phase, extrapolation of 
$E_m(N)-E_{\sigma}(N)$ yields $E_m(x)$. As seen in Fig. \ref{fig4}, $E_m(N) - E_{\sigma}(N)$ 
converges well to $E_m(x)$ on the $x_1$ side and less well on the $x_2$ side where even and 
odd $N/2$ appear for $x_2(N)$ in Table \ref{tb1}.
The largest magnetic gap is $0.29 J$ at $x = 0.40$, close to the $\epsilon_0(x)$ peak in Fig. \ref{fig1} 
and clearly beyond $x_{MG} = 1/3$. The $E_m$ maximum and position agree well with DMRG 
in Fig. 5 of ref. 14. Since Fig. \ref{fig3} shows $E_{\sigma}(x)$ to be very 
small between $x = 0.3$ and 0.5, large $E_m$ in this interval is consistent with 
small finite-size corrections. More accurate DMRG is needed \cite{r25p} for 
$E_m(x)$ at $x > 0.5$.

 To obtain the BOW amplitude $B(x)$, we break inversion symmetry according to
\begin{eqnarray}
H(x,\delta)=H(x)+\delta \sum_n (-1)^n \vec {s}_n.\vec {s}_{n+1}
\label{eq5}
\end{eqnarray}

\begin{figure}[h]
\begin {center}
\hspace*{-0cm}{\includegraphics[width=7.0cm,height=9.0cm,angle=-90]{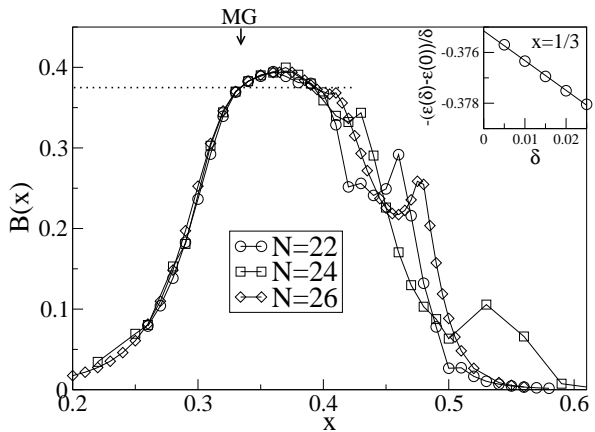}} \\
\caption{Amplitude $B(x)$ in Eq. \ref{eq4} of the BOW of $H(x)$ for finite $N$ and PBC in the BOW phase. 
$B(1/3) = 3/8$ is exact for either $|K1 \rangle$ or $|K2 \rangle$ in Eq. \ref{eq2}. Finite-size effects 
around $x \approx 1/2$ are due to reversals of the ground state’s inversion symmetry. The inset 
shows $-[\epsilon_0(1/3,\delta) - \epsilon_0(1/3)]/\delta$ vs $\delta$ at $N = 20$ for the symmetry-breaking 
perturbation in Eq. \ref{eq5}.} 
\label{fig5}
\end {center}
\end{figure}

\noindent At the MG point, $|K1 \rangle$ is the gs for $\delta = 0+$ and $|K2 \rangle$ for $ \delta = 0-$. The gs energy 
per site, $\epsilon_0(x,\delta)$, gives $B(x) = -(\partial \epsilon_0/\partial \delta)_0$. 
The inset of Fig. \ref{fig5} shows $-(\epsilon(1/3,\delta)-\epsilon(1/3,0))/\delta$ for $N = 20$ 
as a function of $\delta$. The intercept is $B(1/3) = 3/8$ while the slope 
is $\chi_d/2$, the harmonic electronic force constant per site for dimerization 
that will be needed in a later study of lattice vibrations. Fig. \ref{fig5} shows $B(x)$ 
in the BOW phase. $B(1/3) = 3/8$ follows directly from $-\langle \vec{s}_n. \vec{s}_{n+1} \rangle=3/4 $ or 0 for paired and 
unpaired neighbors, respectively. The $N$ dependence of $B(x)$ is negligible 
near the MG point up to $x \approx 0.45$, but it becomes significant around $x \approx 0.5$ 
where the location of the gs crossings in Fig. \ref{fig3} depend on $N$. Finite-size 
effects also appear near $x_1$ and $x_2$ where $B(x)$ becomes small but does not 
vanish. Since $\psi_0(x',\pm)$ with $\sigma = \pm 1$ are degenerate at crossings $x'$, 
the linear combinations $(\psi_0(x',+) \pm \psi_0 (x',-)) /{\sqrt 2} $ are 
broken-symmetry states whose expectation value in Eq. \ref{eq4} leads to

\begin{eqnarray}
B(x')=|\langle \psi_0(x',+)|\sum_n (-1)^n\vec {s}_n.\vec{s}_{n+1}|\psi_0(x',-) \rangle|/N
\label{eq7}
\end{eqnarray}

\noindent The matrix element agrees quantitatively with $B(x') =-(\partial \epsilon_0(x',\delta)/\partial \delta)_0$, as it must. 
The two determinations of $B(x)$ are the same within our numerical accuracy when $E_{\sigma} < 0.01J$.

To our surprise, $B(0.35)$ is slightly but distinctly larger than 3/8, the amplitude at 
the MG point for $|K1 \rangle$ or $|K2 \rangle$. A Kekul$\rm \acute{e}$ diagram has 
perfect AF correlation with one neighbor, which seems to be the limiting case of a BOW. 
While the AF correlation or bond order decreases slightly in systems with $B(x) > 3/8$, 
there is now small F correlation or negative bond order with the other neighbor. 
Direct solution of systems with OBC and $B(x) > 3/8$ yields large positive and small 
negative bond orders that alternate along the chain. We recall that the second-neighbor bond orders, 
$-\langle \vec{s}_n. \vec{s}_{n+2} \rangle$,  are negative for $x < 1/3$, vanish at $x = 1/3$, 
and are positive for $x > 1/3$. The BOW phase for $x > 1/3$ has AF correlations for second 
neighbors and alternating strong AF and weak F correlations for first neighbors. We also note 
that $B(x)$ and $E_m(x)$ are not simply proportional to each other. The $B(x)$ maximum in Fig. \ref{fig5} is at decisively 
lower $x$ than the $E_m(x)$ maximum in Fig. \ref{fig4}. White and Affleck \cite{r13p} 
were also surprised that $d = 2B$ could exceed 3/4 at $J_2/J_1 > 1/2$ and interpreted the 
result as ferromagnetic correlation; Fig. 8 of ref. 14 agrees quantitatively with Fig. \ref{fig5} 
for $0.3 < x < 0.5$.\\

We consider next the excited states of $H(x)$ and present results at 
$x_{MG}=1/3$  that are representative for the interval $0.3<x<0.5$ in which 
$B(x)$ and $E_m(x)$ are large. The gs energy per site at the MG point is 
$\epsilon_0(1/3) = -1/4$ for either PBC or OBC. Table \ref{tb2} list excitations 
with increasing energy for $N =$ 24, 26 and 28. 
Sparse matrix methods \cite{r19,r20} are used for a few states in each symmetry subspace. It becomes 
progressively more difficult numerically to go beyond 3 or 4 states for large $N$. 
The notation $Sr$ indicates total spin S and state index, $r = 1, 2, 3...$. States 
are doubly degenerate with wavevector $\pm k$ except for $k = 0 (\sigma=+1)$ 
and $ \pi (\sigma=-1)$. The lowest triplet at $E_m$ and singlet at $E_3$ decrease slowly with $N$ 
and are known rigorously to be finite in the infinite chain \cite{r10}. Finite-size effects 
are more pronounced with increasing $r$. The gap $E_3- E_m$ decreases with $N$. We expect 
it to vanish in the extended system whose elementary excitation are spin solitons, 
each with $s = 1/2$, with paired or parallel spins. There are additional singlets and triplets below 
the lowest quintet at $E_Q(N)$.

\begin{table}
\begin{center}
\caption {Excitation energies $E(Sr)$ of $H(1/3)$, in units of $J$, for $N$ sites and PBC.}
\begin{tabular}{cccc} \hline
Spin and State, $Sr$ &  N=28 & N=26 & N=24\\\hline
Triplet, T1&  0.1691  & 0.1705  & 0.1727\\
Singlet, S3&  0.1757  & 0.1793  & 0.1839\\
T2, T3     &  0.1833  & 0.1873  & 0.1921\\
S4, S5     &  0.1898  & 0.1953  & 0.2022\\
T4         &  0.2089  & 0.2155  & 0.2242\\
Quintet, Q1&  0.4200  & 0.4324  & 0.4499\\\hline
\end{tabular}
\label{tb2}
\end{center}
\end{table}

\begin{figure}[h]
\begin {center}
\hspace*{-0cm}{\includegraphics[width=7.0cm,height=9.0cm,angle=-90]{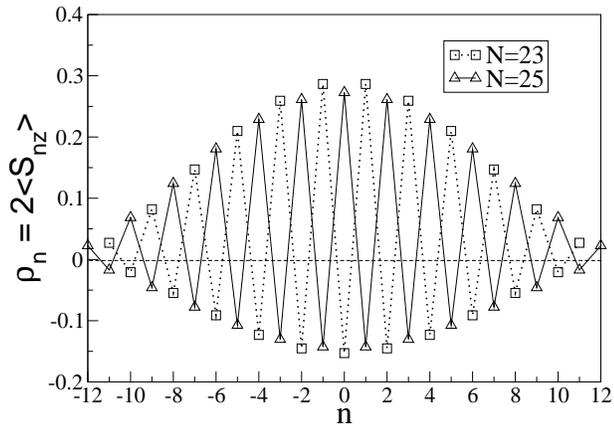}} \\
\caption{ Spin densities $\rho_n = 2 \langle S^z_n \rangle$ in the doublet
 ground state of $H(1/3)$ with OBC and $N$ = 23 and 25. The central site 
at $n = 0$ has large positive $\rho_0$ for $N= 25$ and small negative $ \rho_0$ 
for N = 23, as discussed in the text. The spin soliton is delocalized over 
the central part in either case.}
\label{fig6p}
\end {center}
\end{figure}

Since all sites of $H(x)$ are equivalent for PBC, it is difficult to discern 
solitons even with exact eigenstates in hand. Fortunately, the gs energy per site 
for OBC and even $N$ is again $\epsilon_0= -1/4$, and the gs $|K1 \rangle$ in Fig. \ref{fig1p} has 
alternating bond orders of $3/4$ and 0 along the chain. We consider $H(1/3)$ 
with OBC and odd $N$, either $N = 4p + 1$ or $N = 4p-1$. The gs is a doublet, $S = S^z = 1/2$, 
with spin density $\rho_n = 2\langle S^z_n \rangle$ at site $n$. With central site at $n = 0$, 
the terminal sites are $\pm 2p$ when $N = 4p+ 1$ and $\pm 2(p - 1)$ when $N = 4p - 1$. 
Linear polyenes or VB diagrams rationalize two distinct series 
when $N$ is finite. The pentyl radical $(N = 5)$ has $\rho_0 > 0$ and 
partial single bonds at the center, while the allyl radical 
$(N =3)$ has $\rho_0<0$ and partial double bonds at the center. Soliton 
spin densities of $H(1/3)$ for $N = 25$ and 23 are shown in 
Fig. \ref{fig6p}. Sites with $ \rho < 0$ indicate electronic correlation \cite{r27p} 
and correspond to nodes in uncorrelated H\"{u}ckel or tight-binding theory. \\

\begin{figure}
\begin {center}
\hspace*{-0cm}{\includegraphics[width=7.0cm,height=9.0cm,angle=-90]{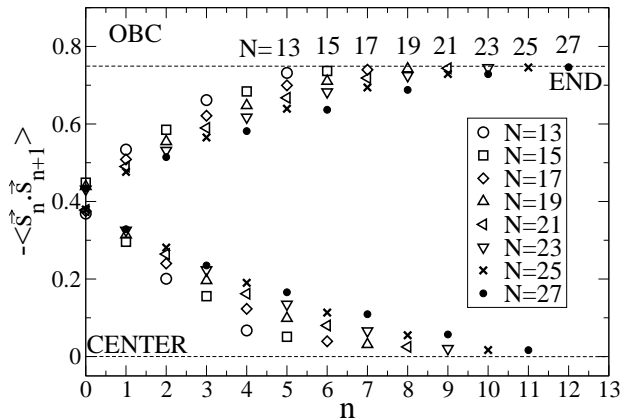}} \\
\caption{Ground-state bond orders, $-\langle \vec {s}_n.\vec{s}_{n+1} \rangle$ of $H(1/3)$, 
for odd N and open boundary conditions (OBC). The bond orders are symmetric about the center, 
$n = 0$, and increase to almost 3/4 at the ends.} 
\label{fig6}
\end {center}
\end{figure}

The gs bond orders of $H(1/3)$ for odd $N$ are close 
to 3/4 at the end and reverse in between. The $(N - 1)/2$ bond orders are 
symmetric about the center, $n = 0$. Fig. \ref{fig6} displays 
$-\langle \vec {s}_n.\vec{s}_{n+1} \rangle$ for different values of $N$. 
Bond orders oscillate with increasing $n$ and grow from the center. 
As expected, the central bond order is slightly larger for the $4p - 1$ series than 
for the $4p + 1$ series. Both spin densities and bond orders are typical of 
spin-1/2 solitons that for $H(1/3)$ connect $|K1 \rangle$ and  $|K2 \rangle$ 
regions. In a BOW phase, broken inversion symmetry and solitons are found in 
regular arrays. Of course, the soliton width $2 \xi$ depends on models and parameters; 
$2\xi$ increases with decreasing dimerization in the SSH model \cite{r13} 
and it also depends on correlations. The results in Figs. \ref{fig6p} and \ref{fig6} 
suggest that spin solitons at the MG point have $2 \xi \approx 15$.\\ 

The energy $2E_W$ of two domain walls is found by comparing the gs energy of even and odd systems 
with OBC and equal length,
\begin{eqnarray}
2E_W(x,N)=2E_0(x,N)-E_0(x,N-1)-E_0(x,N+1)
\label{eq8}
\end{eqnarray}

\noindent At the MG point, we find $2E_W$= 0.1701, 0.1684 and 0.1669 for $N=$23, 25 and  27, respectively,  
slightly less than the $E_m$ values in Table \ref{tb2} for even $N$. $2E_W(1/3,N)$ has weaker $N$ 
dependence than $E_m(1/3,N)$, and a joint extrapolation returns $2E_W = E_m = 0.151$ for the 
infinite chain. Finite-size effects are larger for $E_3(1/3,N)$ in Table \ref{tb2} and even 
larger for $E_Q(1/3,N)$.\\ 

Direct solution up to $N\approx 30 $ indicates that the elementary excitations of 
$H(1/3)$ are spin-1/2 solitions with $2E_w=E_m=E_3=E_Q/2$ in the infinite chain. 
Finite $N$ results suffice for $0.3<x<0.5$ when $B(x)$ and $E_m(x)$ are large. 
Longer chains can be studied using DMRG methods that will be needed for the 
BOW phase of Hubbard-type models. Since 1D systems at $T > 0$ cannot have long-range order, 
topological solitons are generic features of systems with a BOW phase. The present discussion 
is limited to a rigid lattice with purely electronic domain walls, but solitons are also 
expected in deformable lattices with linear electron- or spin-phonon coupling.
  
\section{ Magnetic susceptibility and specific heat}

Static magnetic susceptibility provides by far the most direct comparison with experiment, 
as amply illustrated \cite{r27,r28} by Heisenberg and other spin chains and by spin-Peierls systems.  
The molar spin susceptibility, $\chi_M(T)$, is an absolute comparison for organic radicals 
with small spin-orbit coupling and $g$ close to 2.00236, the free-electron value. Since $H(x)$ 
conserves $S$, the energy level $E_{Sr}$ splits into $2S+1$ Zeeman levels in an applied field. 
The full energy spectrum of $H(x)$ in zero field is required to construct the partition function

\begin{eqnarray}
Q_N=\sum^{N/2}_{S=0} \sum_r (2S+1) {\rm exp}(-E_{Sr}/k_BT)
\label{eq9}
\end{eqnarray}

\noindent  where $k_B$ is the Boltzmann constant and $E_{Sr}$ is excitation energy from the singlet gs 
to the $r^{th}$ energy level with spin $S$.  The molar spin susceptibility of an $N$-site 
chain is \cite{r21}

\begin{eqnarray}
\chi_M(T,N)&=&\frac{N_Ag^2\mu^2_B}{3k_BTNQ_N} \sum^{N/2}_{S=0} \sum_r S(S+1)\times \nonum \\
 & &(2S+1){\rm exp}(-E_{Sr}/k_BT) 
\label{eq10}
\end{eqnarray}
\noindent where $N_A$ is Avogadro's number and $\mu_B$ is the Bohr magneton. 
Finite-size effects become severe when $k_BT$ is small compared to $E_m(N)$. 

\begin{figure}
\begin {center}
\hspace*{-0cm}{\includegraphics[width=7.0cm,height=9.0cm,angle=-90]{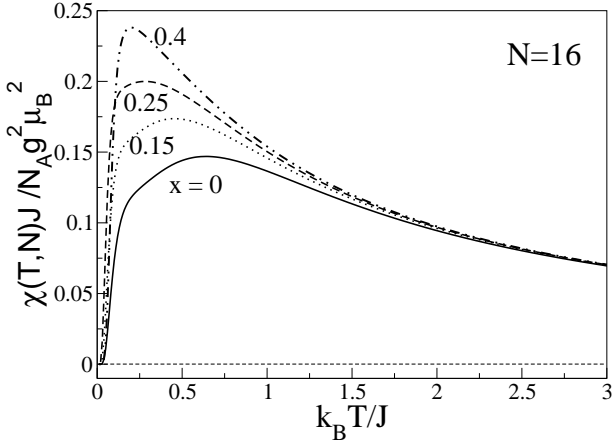}} \\
\caption{Molar spin susceptibility, $\chi_M(T,N)$ in Eq. (10), vs $k_BT/J$ for spin chains with $N = 16$ and  
PBC for $x = 0$, 0.15, 0.25 and 0.40 in Eq. (1). The fluid phase up to $x_1 = 0.1943$ has finite $\chi_M(0)$. 
The curves become independent of $x$ at high $T$.} 
\label{fig7}
\end {center}
\end{figure}

Fig. \ref{fig7} shows $\chi_M$ for N = 16 as a function of 
$k_BT/J$ for several values of $x$. The curves converge 
for $T > J/k_B$ because the total number of spins is the same and so is the Weiss constant $J/2k_B$. The 
number of spins and Weiss constant are the $T^{-1}$ and $T^{-2}$ terms, respectively, at high $T$. 
The $\chi_M(T)$ maxima are well converged at $N =16$, as can be shown by solving $N = 18$ or 14. 
The situation is different as $T \rightarrow 0$, where $x < x_1$ leads to finite $\chi_M(0)$ in 
the fluid phase \cite{r8} while $x > x_1$ has $\chi_M(0) = 0$ due to finite $E_m$. The $x = 0$ 
and 0.15 curves for finite $N$ are dominated by finite-size effects at low enough $T$. The $x = 0.40$ 
curve is almost quantitative since $E_m(0.4)$ exceeds zero-field effects at $N = 16$. The $x = 0.25$ 
curve is intermediate since $E_m$ is finite but $N$ dependent.

\begin{figure}
\begin {center}
\hspace*{-0cm}{\includegraphics[width=7.0cm,height=9.0cm,angle=-90]{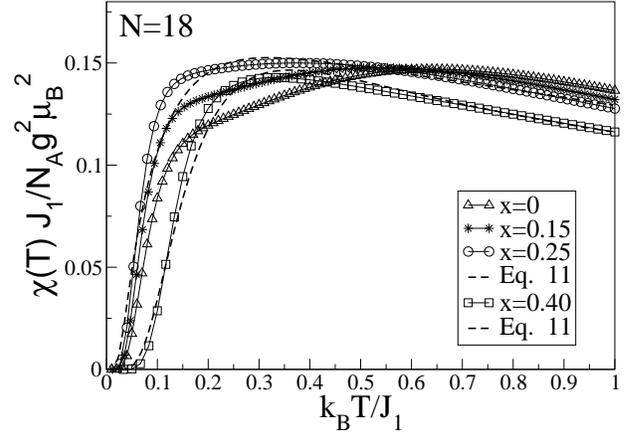}} \\
\caption{Molar susceptibility, $\chi_M(T,N)$ in Eq. \ref{eq10}, vs $k_BT/J_1$, 
the nearest-neighbor exchange, for spin chains with $N = 18$ and  PBC. The dashed lines 
for $x = 0.25$ and 0.40 are Eq. \ref{eq12} with parameter shown in Fig. \ref{fig9}.} 
\label{fig8}
\end {center}
\end{figure}

\begin{figure}[h]
\begin {center}
\hspace*{-0cm}{\includegraphics[width=7.0cm,height=9.0cm,angle=-90]{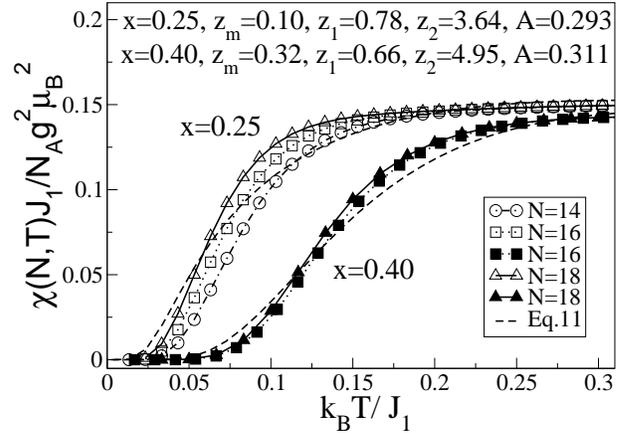}} \\
\caption{Finite-size effects on $\chi_M(T,N)$ in the BOW phase of $H(x)$ at $x = 0.25$ and 
0.40. Large $E_m(0.40)$ gives small changes for $N = 16$ and 18, while small $E_m(0.25)$ 
leads to stronger $N$ dependence. The dashed lines for $x = 0.25$ and 0.40 are 
Eq. \ref{eq12} with the indicated parameters $z_i$ and $A$.} 
\label{fig9}
\end {center}
\end{figure}

For reasons given in the Discussion, we are interested in variable $J_2 = xJ$ at constant $J_1 = J(1-x)$. 
The $\chi_M(T)J_1$ maxima in Fig. \ref{fig8} depend weakly on $J_2$ up to $x = 0.40$; the curves now 
cross because the Weiss constant varies with $x$. The HAF $(x = 0)$ maximum broadens and shifts to 
lower $T$ with increasing $J_2/J_1=x/(1 - x)$. Size convergence at low $T$ is shown in Fig. \ref{fig9} 
for $x = 0.25 (J_2/J_1 = 1/3)$ and $x = 0.40~(J_2/J_1 = 2/3)$. Large $E_m$ at $x = 0.40$ gives convergence at $N$=16 and 18. Small $E_m$ at $x = 0.25$ limits convergence to the broad $\chi_M(T)$ 
maximum. In either case, finite $E_m$ ensures that $\chi_M(0) = 0$ and gives a substantial range in which 
$\chi_M(T)$ is almost linear in $T$. The slope of the linear regime depends weakly on $N$. 

There are many realizations of dimerized HAFs with $x = 0$ in Eq. \ref{eq1} and alternating $J(1 \pm \delta )$ 
along the spin chain \cite{r27,r29}. Dimerized chains with $\delta > 0.3$ are well approximated as $N/2$ 
singlet-triplet (ST) pairs with spin-wave dispersion \cite{r29,r30}. An ST approximation also provides 
insight into the BOW phase. A normalized density $n(E)$ of two-level systems with ST gap $E$ leads to

\begin{eqnarray}
\chi_{ST}(T)=\frac{N_Ag^2\mu^2_B}{k_BT} \int^{\infty}_0 \frac{n(E)dE}{\bigg( 3+{\rm exp}(E/k_BT)\bigg)}
\label{eq11}
\end{eqnarray}

\noindent The integral can be evaluated for any piecewise constant $n(E)$. We consider $n(E) = 0$ aside from two 
intervals; $n(E) = A/(E_1-E_m)$ for $E_m \le E \le E_1$ and $n(E) = (1 - A)/(E_2 - E_1)$ for $E_1 \le E \le E_2$. The molar susceptibility is

\begin{eqnarray}
\frac{\chi_{ST}(T,N)J_1}{N_Ag^2\mu^2_B}&=&\frac{A}{3(z_1-z_m)}{\rm ln}\bigg(\frac{1+3{\rm exp}(-z_mJ_1k_BT)}{1+3{\rm exp}(-z_1J_1k_BT)}\bigg)  \nonum \\
&+&\frac{1-A}{3(z_1-z_2)} \nonum \\
&\times &{\rm ln}\bigg(\frac{1+3{\rm exp}(-z_1J_1k_BT)}{1+3{\rm exp}(-z_2J_1k_BT)}\bigg) 
\label{eq12}
\end{eqnarray}

\noindent with $z_m = E_mJ_1$, $z_1 = E_1J_1$ and $z_2 = E_2J_1$. As expected, the ST gap gives an exponential 
$\chi_{ST}$ at sufficiently low $T$, thereby fixing $z_m$. The width of $n(E)$ is controlled by $z_2$ and 
is fixed by the $\chi_M(T)$ maximum. The shape of $\chi_{ST}(T)$ can be varied by $A$ and $z_1$, 
or by additional parameters when $n(E)$ has more than two intervals. The dashed lines 
in Figs. \ref{fig8} and \ref{fig9} are $\chi_{ST}(T)$ with the parameters in Fig \ref{fig9}. 
The spin susceptibility in the BOW phase is reasonably well modeled with a distribution $n(E)$ of 
ST gaps that is constant in two intervals.\\

The molar specific heat of $H(x)$ for $N$ sites is
\begin{eqnarray}
\frac{C(T)}{N_Ak_B}=\frac{1}{N}\bigg(\frac{J}{k_BT}\bigg)^2(\langle E(T)^2 \rangle-\langle E(T) \rangle^2 )
\label{eq13}
\end{eqnarray}

\noindent  The thermal averages require the energy spectrum 
$E_{Sr}$ and degeneracy. The results below are for PBC. Since 
$S$ is conserved, separate contributions to $\langle E(T) \rangle$ can 
readily be identified. But the entropy is not 
additive in $S$ and there is no unique partitioning of $C(T)dT = TdS$ 
into contributions in $S$. One choice is the temperature derivative of 
the S component of $\langle E(T) \rangle$. Another choice is based on 
fluctuations,

\begin{eqnarray}
\frac{C(T)}{N_Ak_B}&=&\frac{1}{NQ_N}\bigg(\frac{J}{k_BT}\bigg)^2 \sum_{Sr}(2S+1) \nonum \\
& &\times (E_{Sr}- \langle E \rangle)^2 {\rm exp}(-E_{Sr}/k_BT)  
\label{eq14}
\end{eqnarray}

 \noindent with $C_S(T)$ given by the sum over $r$ for fixed $S$. The $C_S(T)$ contributions in 
Eq. \ref{eq14} are manifestly positive and are shown below. We also decomposed $C(T)$ 
based on $\langle E(T) \rangle$. The results are similar at low $T$, the region of 
interest. \\ 

\begin{figure}[h]
\begin {center}
\hspace*{-0cm}{\includegraphics[width=7.0cm,height=9.0cm,angle=-90]{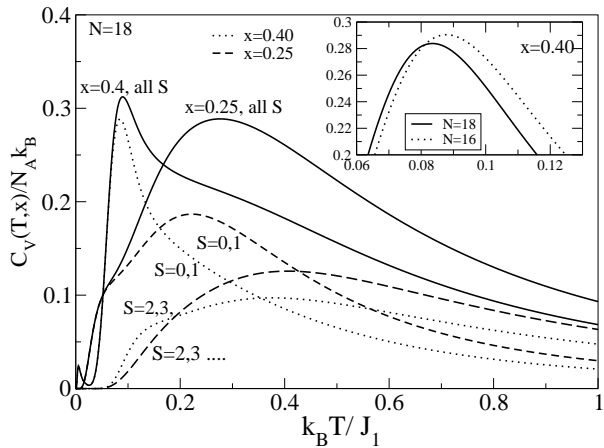}} \\
\caption{Molar specific heat $C(T)$ in Eq. \ref{eq14} in the BOW phase of 
$H(x)$ at $x = 0.25$ and 0.40 for $N = 18$ and PBC. The dotted and dashed lines are 
contributions in $S = 0$,1 and $S \ge 2$, respectively, according to 
Eq. \ref{eq14}. The inset shows the $S = 0$,1 contribution for $x = 0.40$ at $N = 16$ and 18.} 
\label{fig10}
\end {center}
\end{figure}

Fig. \ref{fig10} shows $C(T,x)$ at $x = 0.25$ and 0.40 as a function of 
$k_BT/J_1$ for $N = 18$. The contributions of $S = 0$, 1 and of $2 \le S \le 9$ 
are indicated with dashed lines. Large $E_m$ at $x = 0.40$ gives a $C(T)$ peak due to $S = 0$, 
1 and a shoulder at higher $T$ for $S \ge 2$ contributions that start with $E_Q \approx 2E_m$. 
Small $E_m$ at $x = 0.25$ gives a single broad $C(T)$ peak whose maximum shifts smoothly to 
higher $k_BT/J_1$ in the fluid phase with $x < x_1$. 
Likewise, there is a single $C(T)$ peak when $E_m$ becomes small for 
$x > 0.5$. Finite $E_m$ in the BOW phase shifts the singlet and triplet part 
of $C(T)$ to low energy and separates them from $S \ge 2$ contributions.\\

M${\rm \ddot{u}}$tter and Wielath \cite{r31} reported $C(T)$ results for 
$8 \le N \le 16$ using a different numerical procedure and without resolving 
contributions in $S$. Their $C(T)$ and $\chi_M(T)$ curves are quite similar to ours, 
but not identical. In particular, their $C(T,1/3)$ develops a shallow minimum at 
$N = 16$ that we do not see at either $N = 16$ or 18. Without proposing an 
explanation for the $C(T,1/3)$ maxima, M${\rm \ddot{u}}$tter and Wielath \cite{r31} 
interpreted the discontinuity of the $C(T,x)$ maximum at the MG point as a 
transition from a ``dimer'' to a ``frustrated'' phase that, moreover, survived in the 
limit of large $N$. On the contrary, our results indicate a single phase for $x> x_1$ and 
we understand the $C(T)$ curves in Fig. \ref{fig10} in terms of 
$S = 0$, 1 and $S \ge 2$ contributions. We turn next to $C(T)$ at large $N = 2n$. The 
fraction of singlets among the $2^{2n}$ spin states is

\begin{eqnarray}
f_0(2n)=\frac{(2n)!2^{-2n}}{n!(n+1)!}\approx \frac{(n+1)^{-\frac{3}{2}}}{\sqrt{\pi}}\frac{e}{(1+\frac{1}{n})^n}
\label{eq15}
\end{eqnarray}

\noindent The second expression follows from Stirling's approximation and is accurate to within a few percent 
for $2n = 16$ or 18. The triplet fraction is $9nf_0/(n + 2)$. Since $f_0$ decreases as $(2/N)^{-3/2}$, 
the $C(T)$ contribution from $S = 0$ and 1 becomes negligible compared to $S \ge 2$ in the thermodynamic limit. 
The inset in Fig. \ref{fig10} compares the $S = 0$, 1 maxima for $N = 16$ and 18. The N = 18 maximum 
is lower as expected for Eq. \ref{eq15}. The DMRG results of Feiguin and White \cite{r22} for $C(T,1/3)$ at $N = 32$ 
and 64 show a small shoulder at low $T$ that they attribute to finite-size effects. 
\section{Discussion}
We have characterized the BOW phase of the linear spin-1/2 chain, $H(x)$ in Eq. \ref{eq1}, 
with frustrated Heisenberg AF exchange $J_1 = J(1-x)$ between neighbors and $J_2 = Jx$ 
between second neighbors. Exact VB methods yield the energies and eigenstates of finite 
systems with periodic or open boundary conditions. The BOW phase for $x>x_1=0.1943$ has a broken 
$C_i$ symmetry and finite magnetic gap $E_m(x)$. Our results are most accurate for $0.3<x<0.5$ where large 
$E_m(x)$ ensures small finite-size corrections. Larger $N$, DMRG or other methods will be needed to characterize 
the BOW phase with small $E_m(x)$ or $B(x)$ at $x>0.5$.\\ 

The BOW phase is particularly simple at $x = 1/3$, the MG point, where the exact gs for PBC is either Kekul${\rm \acute{e}}$ 
diagram $|K1\rangle$ or $|K2 \rangle$ in Fig. \ref{fig1p}, and $|K1 \rangle$ for OBC. The BOW amplitude is $ B(x)$ 
in Eq. \ref{eq4}, with $B(1/3) = 3/8$ and a broad maximum in Fig. \ref{fig5} at lower $x$ than the
 $E_m(x)$ maximum in Fig. \ref{fig4}. Topological spin-1/2 solitons that reverse the bond 
order are the elementary excitations of the BOW phase, as shown in 
Fig. \ref{fig6p} and \ref{fig6} at $x = 1/3$ for odd $N$ and OBC. The energy $2E_W$ of two solitons 
corresponds for finite $N$ to $E_m$ for parallel spins or to $E_3$ for paired spins. 
We have also found the consequences of finite $E_m(x)$ on the molar spin susceptibility $\chi_M(T)$ 
and specific heat $C(T)$.\\

The magnetic properties of the EHM with parameters $U$, $V$ and $t$ are closely related to $H(x)$ 
when $t \gg (U -V)$. Van Dongen \cite{r34} mapped the EHM in the spin sector to $H(x)$ with
\begin{eqnarray}
J(1-x)\equiv J_1&=&\frac{4t^2}{U-V}+4J_2 \nonum \\
xJ\equiv J_2&=&\frac{4t^4}{(U-V)^3}  
\label{eq16}
\end{eqnarray}
\noindent The HAF is the familiar limit $t \ll (U - V)$ leading to $x = 0$. 
Increasing $V$ at constant $t$ and $U$ amounts 
to increasing $J_2/J_1 = x/(1 - x)$. The CDW transition of the EHM is close to $V = U/2$. 
Since a continuous CDW transition \cite{r3,r4} requires $t > U/7$, the $t \ll (U - V)$ approximation fails at 
the BOW boundary of the EHM. The next term goes as $t^6$ and in addition to $J_1$, $J_2$ contributions, 
it adds \cite{r35} a four-spin contribution that requires going beyond $H(x)$. Charge degrees of freedom 
cannot be neglected in the BOW phase of the EHM or of related models with Coulomb interactions. 

It is nevertheless attractive to approximate magnetic properties of Hubbard-type BOWs with $H(x)$, 
much as HAFs have been used for Hubbard models. Finite $E_m$ implies an exponentially small $\chi_M(T)$ 
at low T followed by a roughly linear increase up to 0.15 in reduced units. As seen Figs. \ref{fig8} and \ref{fig9}, 
the $\chi_M$ maximum depends weakly on $J_2$. Linear $\chi_M(T)$ vs $T$ behavior is distinctly different 
from an HAF \cite{r30} or a Hubbard model \cite{r36} with $E_m = 0$ and finite $\chi_M(0)$ that exceeds $60\%$ 
of the $\chi_M$ maximum. Linear $\chi_M(T)$ following an onset has been observed \cite{r37} in several 
alkali-TCNQ salts up to $T \approx 450 K$, the limit of their thermal stability. In our opinion, such $\chi_M(T)$ 
in regular arrays are signatures of BOW phases with $E_m > 0$ in Hubbard-type models as well as in $H(x)$.

There are several reasons for considering 
1:1 alkali-TCNQ salts as possible BOW systems. The strongest case \cite{r33} is for 
Rb-TCNQ(II): its $100 K$ structure has regular stacks of $\rm TCNQ^-$ at inversion 
centers, negligible $\chi_M(T)$ us to $150 K$ and infrared spectra that indicate 
broken electronic inversion symmetry. Hubbard-type models have long been used for 
the magnetic, optical and electrical properties of quasi-1D organic ion-radical 
crystals \cite{r15,r29}. The singly occupied MOs of $\rm TCNQ^-$ form a half-filled band. 
The BOW phase of Hubbard-type models is narrow, close to the CDW transition. 1:1 
alkali-TCNQ crystals are close \cite{r38} to the CDW transition based on their 
electrostatic (Madelung) energy and the electronic structure of $\rm TCNQ^-$. 
The spin susceptibility of $H(x)$ is encouraging for a BOW interpretation. More 
quantitative modeling will require values for $t$, $U$, $V$ and other microscopic parameters.\\

In summary, we have characterized the BOW phase of the linear 
spin-1/2 chain $H(x)$ with frustrated first and second neighbor exchange. We exploited the exact 
gs at $x = 1/3$ for finite $N$ to obtain the BOW amplitude $B(1/3) = 3/8$, the magnetic gap $E_m$, 
the spectrum of low-energy excitations, and topological spin solitons between BOWs with 
opposite phases. The spin chain makes possible a detailed examination of a BOW phase. While 
quantitative results are limited to $H(x)$, the consequences of broken inversion symmetry, gs degeneracy and 
finite $E_m$ hold for BOW phases in general. The spin susceptibility of $H(x)$ is consistent with 
the unusual $\chi_M(T)$ of alkali-TCNQ salts with regular stacks and provides additional support 
for the hypothesis \cite{r33} that these salts are physical realizations of BOW phases.
 
{\bf Acknowledgements}. We thank D. Sen for valuable comments concerning the 
BOW phase at large $J_2/J_1$. ZGS thanks A. Girlando for access to unpublished 
RbTCNQ(II) spectra and A. Painelli for discussions about BOW systems. 
Princeton research was supported in part by the National Science Foundations 
under the MRSEC program (DMR-0819860). SR thanks DST India for
funding through SR/S1/IC-08/2008 and JC Bose fellowship.

\end{document}